# Opportunistic and Context-aware Affect Sensing on Smartphones: The Concept, Challenges and Opportunities.

Rajib Rana[*,^]   Margee Hume[♯]   John Reilly[+]   Raja Jurdak[±,x]   Jeffrey Soar[^]

[^]University of Southern Queensland, Australia
[♯]Central Queensland University, Australia
[+]Townsville Hospital Mental Health Service Group, Australia
[±]Distributed Sensing Systems, CSIRO, Australia
[x]University of Queensland, Australia

*Abstract*

*Opportunistic affect sensing offers unprecedented potential for capturing spontaneous affect, eliminating biases inherent in the controlled setting. Facial expression and voice are two major affective displays, however most affect sensing systems on smartphone avoid them due to extensive power requirements. Encouragingly, due to the recent advent of low-power DSP (Digital Signal Processing) co-processor and GPU (Graphics Processing Unit) technology, audio and video sensing are becoming more feasible on smartphone. To utilize opportunistically captured facial expression and voice, gathering contextual information about the dynamic audio-visual stimuli is also important. This paper discusses recent advances of affect sensing on the smartphone and identifies the key barriers and potential solutions for implementing opportunistic and context-aware affect sensing on smartphone platforms. In addition to exploring the technical challenges (privacy, battery life and robust algorithms), the challenges of recruiting and retention of mental health patients have also been considered; as experimentation with mental health patients is difficult but crucial to showcase the importance/effectiveness of the smartphone centred affect sensing technology.*

## Introduction

Automatic affect sensing offers unprecedented opportunity for meaningful human computer interactions[1]. Due to the advancements in smartphone sensor technology, and our extended association with smartphones, affect sensing is feasible on the smartphone platform. Affective information is very useful for wide range of applications. A mental health application can trigger intervention when it detects declining affective states. A music or video recommendation system can make mode-appropriate recommendations and so on.

Existing smartphone affect sensing systems have utilized a large range of sensors, but cameras and microphones are seldom utilized due to their power intensive operations[2]. Therefore, most affect sensing systems have not considered facial expressions and voice; although these are the most important affective displays[1, 3, 4]. Encouragingly, the recent advent of low-power DSP co-processors and GPU is enabling audio[5] and video[6] sensing.

In this paper we focus on affect sensing on smartphones using facial expression and voice. In particular, we propose the concept of opportunistic and context-aware sensing of these two modalities. Opportunistic affect sensing captures spontaneous affect, which reduces the significant bias impacting affect recognition in controlled environements[7]. We also propose capturing contextual information about the dynamic audio and video stimuli, since affective behavior is largely context-dependent[8]. This contextual information will assist users, including clinicians to cross validate affective response against the audio-video stimuli.

Automated affect sensing, including analysis of spontaneous facial expressions and vocal expressions, is much studied, however smartphone affect sensing is relatively new. Existing reviews do not incorporate affect sensing using facial expressions and voice. Lane et al.[9] conducted a comprehensive review of mobile phone sensing, which encapsulates opportunistic and context-aware sensing, but our key focus is on affect sensing, which is different to them. Similarly, Koáakowska[10] reviewed studies, which were only based on phone interactions.

This paper discusses affect and affect models and provides a comprehensive review of affect sensing on smartphones and all data modalities considered to date, including phone usage, phone interaction, sensor data feeds, and proactive reporting. It highlights gaps in affect sensing from facial expression and voice on smartphone and presents the concept of opportunistic and context-aware affect sensing for these two modalities. It further discusses key technical challenges: privacy, battery life and robust algorithms and validation related challenges: patient recruitment and retention, of opportunistic and context-aware affect sensing. We mainly consider mental health patients as this group can be highly benefited from an affect sensing application. It can facilitate self-monitoring of mental health as well as can facilitate informed consultation by making historical affective data available. While addressing technical challenges is necessary for accurate, reliable and secure operations, addressing the validation challenges is crucial for demonstrating the usefulness of the technology.

## Affect Sensing on Smartphones

### Affect

Affect is an observable expression of emotion communicated through affective displays such as facial expressions, gestures, tone of voice, and other signs of emotion such as laughter or tears. Affect is person-dependent[11], as individuals may respond differently to social situations or stimuli. Ekman[12] and Circumplex[13] mood models are the most popular affective models. Ekman model[12] describes affect in terms of six basic emotions: happiness, fear, sadness, anger, disgust and surprise, whereas Circumplex model describes all possible emotions in two-dimensions of arousal and valence.

**Affect Sensing**

Affect sensing on smartphones is relatively new. Based on the sensing modalities, we group the existing studies into five categories: (1) phone usage (2) phone interaction (3) sensor feed (4) hybrid and (5) self-reporting.

**Phone Usage**: includes communication patterns, i.e., statistics of call and text messages. It also includes the frequency of using social media, application usage patterns (for security reason no longer available to a non-system app), and switching between applications.

MoodScope[2] authors showed that "communication history" and "categorized application usage logs" are highly correlated to both dimensions of the Circumplex mood model. Bauer et al.[14] experimented with students during exam (stressful) and non-exam (non-stressful) season and reported that a behavior modification could be clearly observed from the phone usage behavior. Bogomolov et al.[15] used call/SMS logs, proximity data, and self-reported surveys about personality traits[16] to develop a predictive model for recognizing happiness.

This category does not use high-density signals and does not require external devices, therefore consumes less power. It also requires relatively fewer intensive privacy measures compared to audio and video centric approach. However, it mainly considers behavioral aspects, which are likely to vary for individuals. Therefore, a research question would be *does a mood inference model based on phone usage need to be individual-based or can a general population-based model be developed?* Moodscope reports a two months training for improving accuracy. A two months waiting-period can potentially create user dissatisfaction, therefore another research question would be, *how to best select the features to minimize the training period?* Or, *how to get the best trade-off of accuracy versus training period?*

**Phone Interaction:** refers to how an individual interacts with the phone, such as typing speed, touch count etc. Lee et al.[17] considered nine different attributes in this modality, spanning backspace key press frequency to typing speed and achieved 67.5% mean classification accuracy for affect recognition. Gao et al[18] investigated the discriminative power of finger-stroke features during a Fruit Ninja gameplay on an iPod for automatically discriminating between four emotional states: Excited, Relaxed, Frustrated, and Bored. Accuracy reached between 69% and 77% for these four emotional states.

This category imposes by far the least privacy risks and computational requirements. It also unleashes the opportunity to discover new markers. But a research question would be *how the overall accuracy can be significantly increased?* Furthermore, similar to phone usage this depends on human behavior, again raising the question, *can a general mood inference model be developed for phone interaction*?

**Sensor Data Feed:** The introduction of sensor-rich smartphone enables a shift towards more direct affect sensing. Doryab et al.[19] collected noise amplitude, location, Wi-Fi SSIDs, light intensity, and movement to determine sleeping and social behavior to detect change in personal behavior of depressed patients. Vhaduri et al.[20] using GPS traces found that major driving events (e.g., stops, braking) increase stress of the driver. To determine stress Alexandratos[21] used smart watch and a heart rate monitor coupled with a smartphone to gather skin conductance and heart rate data, respectively. Gaggioli et al.[22] used external ECG and accelerometer sensor data to infer stress levels.

This category utilizes many on-board and external sensors, many of which (e.g., the GPS receiver) are highly power consuming. Power saving strategies need to be adopted for reliable operations, which none of the studies in this category have utilized. Contextual information (e.g., an increase in heart rate is due to physical exercise or due to a physiological problems) is crucial to properly characterize sensor data. However, automatic sensing context detection is still in infancy and needs to be developed to meaningfully use sensor data.

**Hybrid Methods:** This category uses a combination of the above three categories. For example, Sano et al.[23] used a combination of sensor feed, phone usage and phone interaction to infer stress level and obtained 75% accuracy in stress detection. Grunerbl et al.[24] also used a combination of sensor feed and phone usage to detect episode changes in bipolar disorder and achieved 76% accuracy in recognizing depressive episode.

This category maximizes the utility of the smartphone by combining phone interaction, phone usage and sensor feed categories. An important research question would be *what is the best way to combine data from low power-consuming phone usage and interaction with the high power-consuming sensor feed data to achieve the best power-accuracy trade-off?*

**Self-Reported/Proactive Affect Sensing:** in this modality, smartphones are used as a vehicle for collecting subjective affective data. Reid et al[25] asked participants to complete a questionnaire asking about locations, activities, companions, mood, stress etc. four times a day. The authors reported that this approach adequately captured the participant's moods, thoughts and activities for 94% of the participants. The questionnaire-based approach of Španiel et al.[26], used SMS-based intervention was used to manage the treatment in early psychosis. Experimenting with 45 patients over 283.3±111.9 days the authors observed a statistically significant 60% decrease in the number of hospitalizations. Yamashita [27] used emoticons to collect subjective mood information and reported a correlation of 0.19~0.88 between mood and emoticons. Jiménez-Serrano[28] also used a questionnaire method to predict postpartum depression and obtained accuracy close to 73%.

While this category avoids inference from sensor data and uses a more direct method to gather affective information, it requires continual, proactive engagement of the user to answer long questionnaire, making it vulnerable to low uptake and low motivation during mental illness. Future research could look into reducing the number of questions or could look into reducing the frequency of completion possibly by incorporating some passive sensing.

# Opportunistic and Context-aware Affect Sensing

**Opportunistic Sensing**

Opportunistic sensing was first introduced as opportunistic people-centric sensing[29], where information related to human activity and their surrounding environment are sensed by portable computational devices carried by individuals in their daily life. In this paper, we use opportunistic sensing specifically for sensing voice and facial expressions. Opportunistic affect sensing helps ubiquitously capture spontaneous affective behavior. The most natural way of opportunistic voice sensing is assessing voice during phone conversations (due to close proximity to the phone microphone). Similarly, a logical way for opportunistic facial expression sensing is assessing facial expression while the user is browsing, writing text messages, playing games or watching movies on the phone, assuming the user is within the view angle of the phone frontal camera.

**Context-aware Sensing**

To date definitions of context are rather broad. Activity, location, ambient sound, ambient light etc. are conventionally used as context. However, our concept of context sensing is specific to voice and facial expressions. We use context to refer to information captured from the smartphone that can relate to voice or facial affect sensing. For facial expressions, context means the type of the content (funny, sad etc.) being browsed, or the type of activity (phone games etc.) the user is involved in. Automatically inferring contextual information will assist cross-validation of affective response against the audio-video stimuli. However, automatically inferring contextual information is challenging[30] and needs much research. One non-intrusive way to analyze the type of the content would be if the user allows listening to the content being watched[31]. For voice context means the type of the conversation (*stressful* e.g., job interview, or *exciting* e.g., a chat with a school friend rediscovering on Facebook). Gathering information about the nature of the conversation is challenging. Generally, it would require a more intrusive approach, such as extracting the content of the conversation. A simpler alternative would be to request user to nominate the type of the conversation at the end of the phone call.

**Landscape of Opportunistic and Context-aware Sensing**

Many studies[32-35] have considered affect sensing from voice during phone conversation or from other day-to-day conversation, but studies that considered affect sensing on phone from facial expression are rare. In StudentLife[35] authors process audio on the fly for audio and speech/conversation detection. They considered the frequency and duration of conversations around a participant as a measure of sociability and used it to determine stress level. VibeFones[32] used tone of voice to obtain an understanding of people's social lives and showed that mobile phones can automatically extract meaning from social and workplace interactions in real-time. SmartMood[33] is a mood tracking and analysis system designed for patients with mania. By analyzing the pitch and amplitude of the voice data captured during phone conversation the onset of new manic episodes was predicted. StressSense[34] is perhaps the most comprehensive work for unobtrusively recognizing stress from human voice using smartphones. Using model adaptation the authors achieve 81% and 76% accuracy for indoor and outdoor environments, respectively. Cho et al.[36] have developed a system for recognizing facial expressions on smartphones, but this does not use opportunistic sensing.

Most of the existing studies have looked into context from a broader perspective. Burns et al.[37] used concurrent phone sensor values from global positioning system, ambient light and recent calls as context for identifying mental health-related states. SmartCoping[38] used location and time of day as contextual data with heart rate variability to infer stress level. Many other studies have considered similar contextual information, but no studies have looked into determining information about the audio-video stimuli causing an affective response.

## Comparison of Existing Literature

In this section we compare (see Table 1) existing literature on smartphone affect sensing. For this comparison we consider a number of important attributes, which are listed below.

**Power Efficiency:** Although the processing capacity of smartphones has significantly increased over the last decade, limited battery power remains a major constraint. Therefore, power efficiency remains an important basis for comparison.

**Population size and Type:** Affect sensing can be used to predict onset of mental illness or stress based on change in affective behavior. It is necessary to experiment with people with mental illness to validate the effectiveness of the underlying prediction models. Experimental size is important to understand the significance of the results.

**Elicitation method:** This will clarify whether the elicitation is posed or spontaneous. Opportunistic sensing is intended to capture spontaneous affective behavior.

**Context:** This is crucial as we offer context-aware affect sensing.

**Accuracy**: is an important candidate for comparison since methods providing high accuracy with small power footprints are more desirable.

**Privacy:** Ensuring privacy is a key requirement to implement opportunistic sensing. We therefore use it as a basis for comparing existing studies.

While the above attributes form the basis for comparing existing works, we propose 3 other attributes, which provide important information about the existing literature.

**Sensing Category:** Phone usage, phone interaction, sensor feed, hybrid method, self-reporting.

**Affect Model/illness:** This important attribute categorizes how a study has used affect sensing e.g. for stress detection or for depression assessment etc. This also indicates which affect models have been used.

**Face/Voice:** This will state whether the study has considered facial expression or voice or both or none.

**Table 1.** Comparison of existing studies [Important attributes are highlighted].

| Ref. | Size; Type | Elicitation Method | Context | Sensing Category | Affect Model/ illness | F/V | Power/ Battery Life | Privacy | Accuracy % |
|---|---|---|---|---|---|---|---|---|---|
| 35 | M:38, F:10; H | Spontaneous | L,T | Hybrid | S, D | Voice | ? | Y | Stress VS activity corrleation:-0.464 |
| 39 | M:3, F:3; H | Posed | A | Feed | S | None | 5h 1500 mAh | N | ? |
| 40 | 5; H | Spontaneous | A, S, L | Hybrid | W | Voice | 15h, 3200 mAh | Y, Audio | 85.3 |
| 41 | M:24, F:11; H | Spontaneous | A, S, L | Hybrid | S | Voice | ? | Y, Audio | 61.5 |
| 42 | M:4, F:10; H | Spontaneous | S | Feed | S | Voice | 9.6h, 1500 mAh | N | 76 Outdoor, 81 Indoor |
| 2 | M:21, F:11, H | Spontaneous | ? | Hybrid | C | None | process: 3.5 mW-hr comm: 1709 mW | Y, Contact ID | Initial: 66 2 months: 93 |
| 43 | M:31, F: 94; H M:4, F: 3; H (Dataset) | Posed | ? | Feed | S, E | Voice | ? | N | Emotion: 75.5 Stress: 93.6 |
| 44 | 4; H | Spontaneous | L,S,T | Feed | B | None | ? | N | ? |
| 18 | M:9, F:6; H | Spontaneous | ? | Interaction | C | None | ? | N | Overall: 77 2 levels of valance:86 |
| 27 | 7; H | Spontaneous | ? | Usage | E | None | ? | N | Correlation: Emoticon VS mood: 0.19~0.88 |
| 15 | 117; H (Dataset) | Spontaneous | W | Usage | E | None | N/A | N | 80.81 |
| 24 | 10; P | Spontaneous | L | Hybrid | B | None | ? | Y, Audio Contact ID | Deteciton: 76 State change: 97 |
| 14 | 7; H | Spontaneous | ? | Hybrid | S | None | ? | N | Behavior change detection accr: 86 |
| 17 | 1; H | Spontaneous | W, L, M | Interaction | E | None | ? | N | 67.52 |
| 33 | M:2, F:2; H | Posed | ? | Feed | M | Voice | ? | N | ? |
| 19 | M:1, F:2; P | Spontaneous | W, T | Hybrid | D | None | ? | N | ? |
| 37 | M:1, F: 7; P | Spontaneous | L | Hybrid | D | None | ? | N | Location prediction accr: 60- 91 |
| 23 | M:15, F:3; H | Spontaneous | ? | Hybrid | S | None | ? | N | 75 |
| 21 | M;12, F:4; H | Posed | ? | Feed | S | None | ? | N | 82.9 |
| 38 | ?; H | Spontaneous | L, M | Feed | S | None | ? | Y, Contact ID | ? |
| 20 | M:15, F:15; H | Spontaneous | ? | Feed | S | None | ? | N | Stress VS GPS trace Correlation: 0.72 |
| 36 | M:65, F:145; H (Dataset) | Posed | ? | Feed | E | Face | ? | N | ? |

[Size; Type: Number of **M**ale/**F**emale participants and Type (**H**ealthy control/Mental health **P**atients). **Dataset** indicates this study has conducted simulations.

Context: **A**ctivity, **L**ocation, ambient **S**ound, ambient ligh**T**, **W**eather, **G**ender, Ti**M**e.

Affect Model/illness: **E**kman[12], **C**ircumplex, **D**epressed, **S**tress, **B**ipolar, **M**anic, **W**ellbeing.

Power/Battery Life: Power Efficiency/Battery life.

Privacy: **Audio**: audio privacy has been applied. **Contact ID**: contact ID has been removed from call logs and SMS.

**?**: missing information.]

**Findings from the comparison**

Comparison of the existing literature (in Table 1) leads to the following key observations:

1. Opportunistic facial expression sensing has not been studied before but opportunistic voice sensing exists to a small extent as two studies have extracted affective information from voice during phone call. This shows the opportunity to develop new methods for opportunistic sensing of face and voice.
2. Relevant context information about dynamic audio-video stimuli has not been extracted before but other contextual information such as weather, ambient light etc. has been utilized for affective inference. Methods for automatically determining such contextual information are therefore warranted, as this will enable automatic cross-validation of affective response to the audio-video stimuli.
3. Only a small number of studies have adopted privacy-preserving measures where the focus is mainly on anonymizing audio and contact ID. Privacy measures for protecting images were not found. Given that privacy is a major hurdle to using technology for this purpose, serious effort needs to be invested to develop privacy-preserving affect sensing and context-extraction strategies.
4. Only one study has engaged a small number of mental health patients. Opportunistic affect sensing can potentially complement current mental health treatment by enabling early prediction of relapse of mental illness. However, experiments need to be conducted with mental health patients to validate the feasibility of having such predictive model.
5. Classification accuracy is poor especially when elicitation is spontaneous. Capturing spontaneous affect is difficult as acquisition cannot be controlled and capture takes place in dynamically changing environment. This highlights the need for developing robust algorithms, which can accurately determine affective information from noisy facial images and noisy voice recordings.
6. A very small number of studies report battery lifetime while using the affect sensing application. But battery usage is a crucial information, since it can potentially determine whether the user is willing to use the application. In the existing studies, phone battery (standard 1500mAh) lasts for around 8-9 hours, which needs to be extended to at least 12 hours, since people tend to recharge their phone overnight during sleep time. Incorporating higher capacity battery can be a potential solution, however this could potentially increase the weight/size of the phone. A more achievable solution may be to utilize low-power co-processors and GPUs to extend battery lifetime.

## Challenges of Opportunistic and Context-aware affect Sensing

Our review shows that opportunistic and context-aware affect sensing is still in its infancy. We discuss in detail the key challenges to making opportunistic and context-aware affect sensing a reality.

**Privacy and Data Security:**

***Privacy and security risks during data acquisition:*** In opportunistic affect sensing privacy can be violated in many ways. Other's conversations might be unintentionally recorded. If shared, the system could gather other's facial expressions and voice recordings. More than one person's face could be present in image, if someone else comes close to the phone while the user sends the text message. Facial images containing a point of interest can reveal presence in those locations[45]. By analyzing characteristic sound patterns (present in the voice recording) that are unique to certain events, location and presence in that event can be determined[45].

Admission control strategies need to be developed to avoid the possible privacy violations. In order to discard voices of additional speakers, one of the techniques could be used is speaker classification algorithms to retain the speech of the dominant speaker (assuming he/she is the phone owner) across all/most recordings and discard other speech samples. Sounds other than speech need to be isolated and discarded to avoid revealing user's presence in certain events. This will require classifying speech and non-speech sounds and discard all non-speech sounds. In several cases speech and background sound could be mixed-up together. Methods/policies need to be developed either to discard those speech samples or to extract speech from those mixed sound samples. Face images need to be searched for multiple face images and any non-dominant (based on frequency of appearance) face image needs to be discarded. To make sure no raw data leave the phone admission control methods need to run on the phone. The key challenge to ensure reliable and secure operation is that the admission control methods need to be strictly privacy preserving while incurring low energy costs.

***Privacy and security risks during data processing:*** One method to ensure data security is to not let any raw information leave the phone and communicating features while performing cloud offloading. However, raw data sometimes can be reconstructed from the features, which are widely used for face and voice classification. For example, the most popular MFCC audio features, which also offer the highest accuracy, can be used to reconstruct original speech[34]. Similarly, Eigenfaces, a popular facial image feature, can be used to reconstruct face images. A research question would be, *how to choose/construct a privacy-preserving feature set to achieve the best privacy versus accuracy trade-off while extracting affective information from speech and face images?* Selection of privacy-aware features can potentially reduce the risks of compromising individual's privacy while the construction of a classification/predictive model using data from multiple people.

**Battery Life:**

Audio and video processing consume substantial amount of power[6, 42] on smartphone. A basic monitoring and classifying of audio on a ARM11 Core processor can consume around 335 mW continuously[42] (comparable to continuously sampling raw GPS data[2]). Similarly a face recognition algorithm on a ARM Cortex A9 CPU consumes as high as 45.9J[6]. Currently, there are two ways to extend battery life: (1) by offloading the processing tasks to a resourceful computer (cloud offloading) or (2) by utilizing the low-power DSP co-processor or GPU. However, both of these have their merits and demerits. For continuous audio classification, continuously powering a wireless connection for constantly uploading audio features and downloading of classification results is extensively power consuming[46]. It will also result in high latency. But cloud processing can take significant processing load from the CPU and could accommodate applications requiring heavy processing. On the other hand DSP co-processor is considerably slower than CPU[5]. It also has serious memory constrains and limited functionality[5]. But careful utilization of DSP co-processor utilization can extend the battery life by 3 to 7 times[5]. Similarly, a case study of GPU-assisted face feature extraction shows a 3.98x reduction in total energy consumption[6]. Further investigations need to be conducted to find out whether cloud offloading and DSP co-processor utilization individually or in combination offer the best accuracy versus energy trade-offs. More importantly, *while used in combination do any of these methods need to be assigned higher weight* (such as, ratio of using cloud-offloading and co-processor processing is 6:4 etc.)? Also, *should these weights be dynamically adapted to the input frequency, i.e., based on the frequency of speech samples arrival (continuous conversation versus intermittent conversation)?*

**Robust Classification Algorithms:**

Many past studies have considered facial expression classification from noisy input images. However, classifying opportunistically captured facial images poses additional problems caused by (1) low image quality (captured by frontal camera), (2) unconstrained environmental conditions (mainly lighting) and (3) rapid motions that occur when the camera is held in the hands[47]. Sparse projections derived from Haar-like features can be used to obtain optimal performance for smartphones as those assure robust classification[47] accuracy. Local binary pattern (LBP) features which can be extracted quickly[48] in a single scan through the raw image are an ideal candidate for smartphone. In addition, LBP features are suitable for low-resolution images[48], which will accommodate the relatively low-resolution photos taken by phone frontal camera. Finally, Sparse Random Classifier (SRC), has been used very successfully for robust (in presence of occlusion, illumination and pose variation) face classification[49]. How to optimize SRC for robust facial expression classification on smartphone requires more investigations.

Although a number of studies (e.g. StressSense, SmartMood and VibeFones) have conducted opportunistic affect sensing, more experiments are needed to validate them in dynamic home environments. Sparse Approximation based Classification[50] methods, which can significantly reduce processing overhead while offering robust audio classification, need to be utilized to obtain high classification accuracy in dynamic audio environments. Given the tremendous success of deep learning in speech recognition, Lane et al.[5] have initiated work for emotion recognition from speech using Deep Learning. Some possible future directions from this work would be, *how to incorporate robustness in Deep Learning for emotion detection in noisy conditions? How to inject sparsity in Deep Learning models for achieving fast processing and robust emotion classification on smartphone?*

**Recruit and Retain Mental Health Patients:**

Opportunistic and context-aware affect sensing can potentially play a significant role in mental health service delivery. However, of the many studies discussed to date, only one[24] has engaged mental health patients in experiments. In order to demonstrate importance of a particular algorithm/method for mental health patients, it is necessary to conduct such experiments with them[33, 51].

The challenge of recruiting and retaining mental health patients is not trivial. In particular, the proposed opportunistic and context-aware smartphone sensing technique could be perceived as invasive by many participants. A possible strategy to gain trust is to visually demonstrate the end-to-end application - from data collection to data processing - to data dissemination; and demonstrate how privacy is preserved and access control is provided in every step. Some participants will only take part in research when they can identify and understand the validity and relevance of the research[52]. Therefore, it is important to demonstrate/project how the proposed research could potentially transform/assist mental health treatment. Some participants might also conduct a personal cost–benefit (travel time and cost) analysis when deciding to participate in research[53]. Offering incentives and flexible meeting times can minimize time, travel, financial and inconvenience costs for participants. Another important consideration is that when deciding the sample size, the researchers also need to take into consideration the possibility of dropout so that results remain statistically valid.  Some additional recruitment would compensate for any future dropouts.

## Conclusion

Opportunistic and context-aware affect sensing is the next generation of sensing on smartphone. It can offer mood inference on the phone, which is valuable for a large range of applications including stress and mental health management. Surveying and analyzing existing literature we identify privacy and data security, patient recruitment, robust computing and battery power as the key challenges for opportunistic and context-aware affect sensing. We have discussed potential solutions to address these key challenges, which will guide researchers to innovate in this space.